\documentclass[a4paper,11pt]{article}
\usepackage{pos}
\usepackage{cleveref}
\usepackage{gensymb}
\usepackage{slashed} 

\title{Predictions for Composite Higgs models from gauge/gravity dualities}

\author*{Werner Porod}

\affiliation{University of W\"urzburg, Institute of Theoretical Physics and Astrophysics \\
        Campus Hubland Nord, Emil-Hilb-Weg 22, D-97074 W\"urzburg, Germany}

\emailAdd{werner.porod@uni-wuerzburg.de}

\abstract{
Gauge/gravity dualities provide a very useful approach into solving strongly coupled systems. We apply this to Composite Higgs models 
and determine the mass hierarchies of the corresponding bound states. As a cross check we
apply this to QCD and compare the results to existing 
lattice calculations for which we find good agreement. We then focus on a particular example whose phenomenology has recently 
been studied in the literature in a generic way and outline first phenomenological implications of our findings for the spectrum.
}

\FullConference{Corfu Summer Institute 2023 "School and Workshops on Elementary Particle Physics and Gravity" (CORFU2023)\\
 23 April - 6 May , and 27 August - 1 October, 2023\\
Corfu, Greece\\}

\begin{document}
\maketitle

\section{Introduction}
The extension of the the AdS/CFT
correspondence \cite{Maldacena:1997re,Witten:1998qj,Gubser:1998bc} to
less symmetric gauge/gravity dualities has proven to be a powerful tool in describing
group  by adding probe branes \cite{Karch:2002sh}. This allows
to  study the related meson operators \cite{Kruczenski:2003be,Erdmenger:2007cm}. 
These methods were successfully used to obtain gravity duals of chiral symmetry
breaking ($\chi SB$) in confining non-abelian gauge theories 
\cite{Babington:2003vm,Kruczenski:2003uq}. 
It is natural  to apply this approach to other 
strongly coupled theories. We focus here on composite Higgs models,
see e.g.~\cite{Panico:2015jxa,Cacciapaglia:2020kgq} for reviews.
These models are characterised by strongly coupled gauge theory and an underlying
set of fermions dubbed hyperquarks in the following.

In composite Higgs models $\chi SB$ in the fermion sector is caused by a 
by a strongly coupled gauge theory, similar to QCD. In this way at least four
Nambu-Goldstone bosons are generated \cite{Dugan:1984hq}. By weakly gauging
part of the global chiral symmetries, four then pseudo-Nambu Goldstone bosons (pNGBs)
can be placed in the fundamental representation of SU(2)$_L$ to become the complex Higgs field.  The composite nature of the Higgs removes the huge level of 
fine tuning in the Standard Model (SM) hierarchy problem. This strong dynamics would 
occur at a scale of a few TeV, the expected mass scale for the bound states. 

We use non-conformal gauge/gravity models that explicitly include the gauge
theories' dynamics through the running of the anomalous dimension $\gamma$ of the 
hyperquark  mass \cite{Erdmenger:2020lvq,Erdmenger:2020flu}. 
The models are inspired by top-down models involving probe D7-branes embedded into 
ten-dimensional supergravity. However, this is combined with a  phenomenological 
approach and sensible guesses for the running of $\gamma$ which are based
on perturbation theory. With this we obtain prediction for part of the mesonic and 
baryonic spectrum of the theory. 
Here we will demonstrate that power of this methods by comparing our results with data for QCD bound states.
Then we will focus on two models: (i) an Sp(4) 
theory with 4 fundamental and 6 sextet Weyl fermions \cite{Barnard:2013zea} for which also lattice
studies exist \cite{Bennett:2019cxd,Bennett:2019jzz}; and an
Sp(4) theory with five sextet Weyl and 6 fundamental fermions
\cite{Ferretti:2016upr}. Further examples can be found in ref.~\cite{Erdmenger:2020flu}.

\section{Gauge/gravity duality, basic idea and some technical aspects} \label{sec: QCD}

We introduce here briefly the model based on gauge/gravity duality, which was first suggested 
in ref.~\cite{Alho:2013dka}. 
In  \cite{Erdmenger:2020flu} it has been dubbed  {\it Dynamic AdS/YM}  
(Anti-de Sitter/Yang-Mills)   to emphasise that it can be used to holographically describe 
the chiral symmetry breaking dynamics of  any gauge theory.
These models include hyperquarks in several, potentially inequivalent, representations. 

We summarize here key elements and refer to \cite{Erdmenger:2020flu} for further 
details. In this modelling, the renormalization group (RG) scale emerges as an extra holographic direction.  
In a five dimensional AdS$_5$ space one has the metric
\begin{align}
ds^2 = \frac{dr^2}{r^2} + r^2 dx_{(1,3)} \,.
\end{align}
The radial direction $r$ is interpreted as the RG scale and slices at some fixed 
$r$ describe the gauge theory on the corresponding $1+3$ dimensional sub-space. 
Fields in the AdS bulk have solutions which describe gauge invariant observables,
i.e.\ operators $\mathcal{O}$ and sources $\mathcal{J}$.

In QCD with two light flavours the chiral flavour symmetry SU(2)$_L \times$SU(2)$_R$ is broken
to SU(2)$_V$ by the formation of a vacuum expectation value  for the operator 
$\bar{q}_L q_R  +$~h.c.
Holographically the operator is described as a dimension one 
scalar field $L$ in AdS$_5$ satisfying the Klein-Gordon equation  \cite{Erdmenger:2007cm}
\begin{align}
\frac{\partial}{\partial r} \left(  r^3 \frac{\partial L}{\partial r} \right)
-r \, \Delta m^2\, L = 0 \,, \quad 
L = \frac{m}{r^\gamma} + \frac{\langle \bar{q} q \rangle}{r^{2-\gamma}}\,,\quad
\gamma (\gamma -2) = \Delta m^2 \,.
\end{align}
The solutions describe a dimension one mass $m$ and the dimension three quark condensate 
$\langle \bar{q} q \rangle$ if $\Delta^2 m = 0$. For non-zero $\Delta m^2$ these 
operators develop an anomalous dimension $\gamma$.

In AdS$_5$ the instability bound for a scalar is given by the Breitenlohner-Freedman 
(BF) bound
\cite{Breitenlohner:1982jf} corresponding to $\Delta m^2 = -1$. The massless $L = 0$ solution becomes unstable
if $\Delta m^2$ passes through this bound. This corresponds to 
$\gamma = 1$ or in other words if
the dimension of $\bar{q} q$ falls to two then there is an instability due 
to the generation of a quark
condensate and $\chi SB$ takes place. Note, that this criteria matches that from gap equation analysis 
in ref.~\cite{Appelquist:1996dq}.
Therefore, one can make holographic models of $\chi SB$ which are 
essentially Higgs-like theories with the scalar field $L$ but whose potential $\Delta m^2$
changes with the RG scale $r$. The potential is determined by the running of the 
anomalous dimension $\gamma$
which in turn is determined by the gauge dynamics. 
Chiral symmetry breaking occurs at the scale where $\gamma = 1$.

Holography allows one to determine the low energy mesonic theory as follows. 
For a given background vacuum, e.g.~a solution for the field $L$, one allows 
small fluctuations of the form $\delta(r)e^{ik\cdot x}$ with $k^2 = -M^2$, which
describe fluctuations of the $\bar{q} q$ operator, for example the $f_0$ state in QCD.
Generically one obtains a  Sturm-Louville system which only has normalizable solutions 
for particular values of $M^2$. Other
states such as spin-1 states like the $\rho$ meson can be included via a gauge field dual 
to $\bar{q}\gamma^\mu q$ and so forth. In this way one obtains the meson spectrum. 
Substituting these solutions back into the
action and integrating over the radial direction $r$ yields a $1+3$ dimensional theory of 
the mesons and their interaction couplings. In a similar way one can also include
baryons and their couplings with the mesons \cite{Erdmenger:2020flu} which
eventually also include the Yukawa coupling of the top-quark \cite{Erdmenger:2020lvq,Erdmenger:2020flu}.

Last be not least, we note, that holography is a strong weak duality and so the 
gravitational dual should only describe the infrared (IR) region of an asymptotically 
free gauge theory with a large coupling. Models based on gauge/gravity duality show some remnant of the 
N=4 super Yang-Mills theory left in the description that enforces conformality in the UV.
This matches the weak coupling physics of these models
and one might be tempted to let the dual extend to the far UV. 
However, one should cut off the description, e.g. around a few GeV in QCD, where the 
coupling becomes weak. Here, one can simply impose a cut off in the bulk but then
there is a matching problem: one should align the dual to meet at QCD the 
intermediate coupling. This means the inclusion of some higher dimension operators (HDO)
at the UV boundary $\Lambda_{UV}$. This can be implemented
using Witten's multi-trace prescription \cite{Witten:2001ua} as has been detailed 
in ref.~\cite{Erdmenger:2020flu} to which refer for further details. The basic
idea behind this is, that one catches in this way the first corrections describing the stringy
nature of excited states. 
In ref.~\cite{Erdmenger:2020flu} the following operators have been considered 
\begin{equation}  
\label{qcdhdo} 
{g^2_S \over \Lambda_{UV}^2} |\bar{q} q|^2\, , \quad
{g^2_V \over \Lambda_{UV}^2} |\bar{q} \gamma^\mu q|^2\, , \quad
{g^2_A \over \Lambda_{UV}^2} |\bar{q} \gamma^\mu \gamma_5 q|^2\, , \quad
{g^2_{\rm B} \over \Lambda_{UV}^5} |q q q|^2\, , 
\end{equation}
with $g_i$ being dimensionless couplings.

We first demonstrate the power of the gauge/gravity duality by applying it to QCD as a test case. 
The corresponding
model is described by the action \cite{Alho:2013dka,Erdmenger:2020flu}
\begin{align}
S = \int d^4 x d\rho \,\text{Tr} \rho^3 \left[
\frac{1}{\rho^2 + |X|^2}  |D X|^2 + \frac{\Delta^2m}{\rho^2} + \frac{1}{2g^2_5}
(F_V^2+F_A^2) \overline{\Psi}(\slashed{D}_{\text{AAdS}} -m) \Psi \right] 
\end{align}
where $X = L e^{2i\pi T}$ describes the quark condensate/$\sigma$ and pion fields, 
$F_V$ the $\rho$ meson and $F_A$ the $a$ meson. $\Psi$ is a Dirac field corresponding
to the nucleon. The factors with $r^2=\rho^2 + |X|^2$ of $X$ in the metric
are deduced from top down (probe D7 brane) models \cite{Erdmenger:2007cm} and are the
simplest ansatz to communicate the background quark condensate to the fluctuation fields.

\begin{figure}[t]
\begin{center}
\includegraphics[scale=0.48]{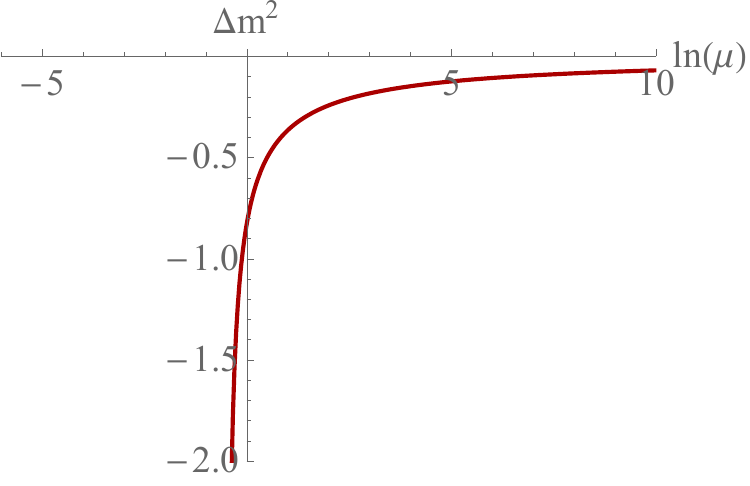} \quad
\includegraphics[scale=0.48]{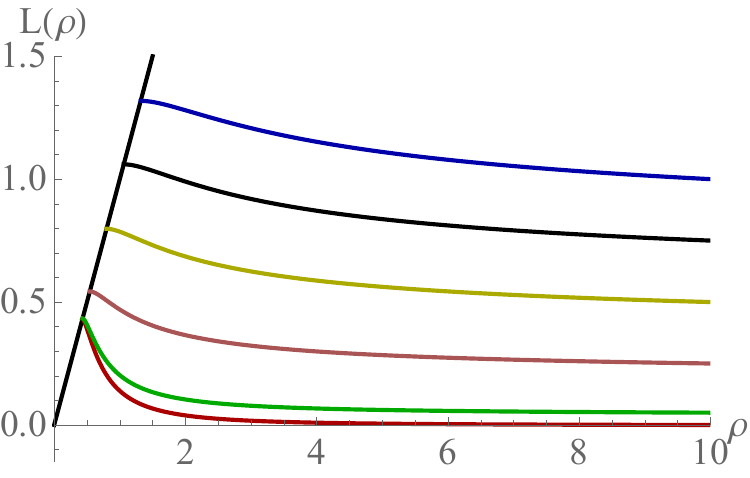} 
\end{center}
\caption{\label{fig:QCD_embedding} The $N_c=3, N_f=2$ QCD model: 
on the left we display the running of the AdS scalar mass $\Delta m^2$ against log RG scale 
(we use $\mu=\sqrt{\rho^2+L^2}$ in the holographic model). On the right we show the the 
vacuum solution for $|X|=L(\rho)$ against  $\rho$.  The 45$\degree$ line is 
where we apply the on mass shell infrared boundary condition. 
The $L(\rho)$ with a massless UV quark has $L_{IR}= 0.43$. The quark masses from top to 
bottom are 1, 0.75, 0.5, 0.25, 0.05, 0. Here units are set by $\alpha(\rho=1)=0.65$.
See ref.~\cite{Erdmenger:2020flu} for further details.}
\end{figure}

The starting point is the perturbative results for the running of $\gamma$. Expanding 
$\Delta m^2 = \gamma (\gamma-2)$ at small $\gamma$ gives
\begin{align}
\Delta m^2 = - 2 \gamma  = - \frac{3({N_c}^2-1)}{2 N_c \pi} \alpha\, 
\end{align}
with $\alpha$ being the QCD coupling constant. Here we allow ourselves 
to extend the perturbative results as a function of renormalization RG scale 
$\mu=r=\sqrt{\rho^2+L^2}$ to the non-perturbative regime.
The resulting running of $\Delta m$ in the Dynamic AdS/QCD
model is shown in \cref{fig:QCD_embedding} on the left -- the BF bound is violated 
close to the scale  $r = \mu = 1$.

In table~\ref{tab:QCDspectrum} we display our results for the spectrum. Note, that we use here and below 
the following notation: we label the model as AdS/SU(3) to indicate the gauge
group and \mbox{2$F$ 2$\bar{F}$} to show that there are 2 Weyl fermions in the fundamental and 
two in the anti-fundamental representation which is equivalent to 2 Dirac fermions in 
the fundamental representation. The second column gives the measured  masses of the
QCD bound states whereas
the third one gives the results for  AdS/SU(3) setting all quark masses to zero. Moreover,
we have used  the $\rho$-meson mass to set the scale. We see, that the ground states
are reasonably well described but the pion decay constant is somewhat low which is partially
caused by neglecting the  UV quark mass(es). The $\sigma$ (S) mass is high, but possibly should 
be compared to the f$_0$(980) if the f$_0$(500) is a pion bound state. The proton mass 
is clearly too high. The first excited states are clearly more off but 
this had to be expected
as no string like structures are taking into account. The situation clearly improves
once HDOs are taken into account as shown in the fourth column. However, this comes
at the cost of fewer predictions at this stage as we have fitted the coefficients of
the HDOs to data because a calculation from first principles is quite difficult.

\begin{table}[t]
\begin{center}
\begin{tabular}{|c|c|cc|cc|}
\hline
   & QCD & AdS/SU(3) &  & AdS/SU(3) & with\\
   &     & 2$F$ 2$\bar{F}$ & & 2 F  2 $\bar{F}$ & HDO \\  \hline
    $M_{\rho}$  &  775  &  775$^*$ &                 & 775$^*$  &    \\
    $M_{A}$     & 1230  & 1183     & - 4\%& 1230$^*$ & $g^2_A=5.76149$ \\
    $M_{S}$     & 500/990 & 973   & +64\%/-2\% & 597  &   $+ 20\%/-40\%$ \\
    $M_{B}$     & 938 & 1451 & +43\% & 938$^*$ &  $g^2_B=25.1558$ \\ \hline
    $f_{\pi}$   & 93 & 55.6  & -50\%  & 93$^*$ &   $g^2_S= 4.58981$ \\
    $f_{\rho}$  & 345 & 321 & - 7\% & 345$^*$  &    $g^2_V=4.64807$ \\
    $f_{A}$     & 433 & 368   & -16\% & 444 &   $+2.5\%$ \\ \hline
    $M_{\rho,n=1}$ & 1465  & 1678 &  +14\% & 1532  &    $+4.5\%$ \\
    $M_{A,n=1}$  & 1655   & 1922   & +19\%  & 1789 &   $+8\%$\\
    $M_{S,n=1}$  & 990 /1200-1500   & 2009  &+64\%/+35\%   & 1449  &   $+46\%/0\%$\\
    $M_{B,n=1}$  & 1440  & 2406  & +50\% & 1529 &   $+6\%$\\ \hline
\end{tabular}
\end{center}
\caption{\label{tab:QCDspectrum} Predictions for masses in MeV for $N_f=2$  SU(3) gauge
theory. The columns show the QCD value  \cite{ParticleDataGroup:2022pth}, the holographic prediction for the theory with
massless quarks and the improved holographic QCD theory with HDOs.  
The $\rho$-meson mass has been used to set the scale and other quantities that 
are used to fix the HDO couplings are indicated by *s.}
\end{table}

\begin{table}[t]
\begin{center}
  \begin{tabular}[h]{|c|c|cc|}
    \hline
     &  QCD  & AdS/SU(3) & \\
     &      &   3 F  3 $\bar{F}$ & \\
    \hline
     $\rho(770)$, $\omega(782)$ & $775.26\pm0.23$ & 775*& \\
    $K^*(892)$ &  $891.67\pm0.26$ & 966& 8\%\\
    $\phi(1020)$ &$1019.461\pm0.016$ & 1120& 9\%\\ \hline
    $a_1(1260)$, $f_1(1260)$&  $1230\pm40$ & 1103 & 11\%\\
    $K_1(1400)$& $1403\pm7$ & 1432 & 2\%\\
    $f_1(1420)$&  $1426.3\pm0.9$ & 1847 & 26\%\\\hline
    $a_0(980)$, $f_0(980)$ &  $980\pm20$ & 930 &5\% \\
    $K_0^*(700)$& $845\pm17$ & 987 & 16\%\\
    $f_0(1370)$ &  1370 & 1031& 28\%\\ \hline
    $\pi^{0,\pm}$ &  $139.57039\pm0.00017$ & 128 & 9\%\\
    $K^{0,\pm}$& $497.611\pm0.013$ &497 & O(0.1) \%\\
    \hline
  \end{tabular}
  \end{center}
\caption{\label{tab:QCD_3flavour} Meson masses in MeV in the three flavour case compared with the 
experimental data \cite{ParticleDataGroup:2022pth}. 
We have fixed the masses for the vector bosons $\rho$ and $\omega$ as indicated by the *
and calculated the masses for the axial vectors, the scalars and the pNGBs. 
  The UV quark masses used are $m_u=m_d=3.1$ MeV and $m_s=95.7$ MeV.  }
\end{table}  

We note here for completeness, that strictly speaking these results do not 
fully take into account the flavour structure of the theory, e.g.\ they do not
take into account for example mass splittings between different isospin multiplets
nor the fact that the quarks have different UV masses. This can be taken into account
by starting with a non-abelian Dirac-Born-Infeld action as has been shown in 
ref.~\cite{Erdmenger:2023hkl}. Table~\ref{tab:QCD_3flavour} gives a corresponding results
from ref.~\cite{Erdmenger:2023hkl}
assuming $m_u=m_d \ne m_s$ and one finds a reasonable agreement between data and our
model.

\section{Example Sp(4) gauge theories}

We focus here on Sp(4) gauge theories which have been proposed in 
refs.~\cite{Barnard:2013zea,Ferretti:2013kya,Ferretti:2016upr,Belyaev:2016ftv} as candidates for CH models.
Moreover, phenomenological aspects of a  particular example of such models, dubbed M5 in ref.~\cite{Belyaev:2016ftv},
have been recently worked out in a series of papers, see \cite{Cacciapaglia:2021uqh,Banerjee:2022xmu,Cacciapaglia:2022bax,Flacke:2023eil,Cacciapaglia:2024wdn}. We will briefly comment on the impact of our findings to phenomenological aspects
in the subsequent section.
We discuss first a realisation that allows for a comparison with lattice data. It
contains 4 Weyl fermions in the fundamental representation $F$. Thus, the
global group containing eventually the electroweak sector of the SM is
SU(4) which gets broken to Sp(4) via the condensate. One can introduce top
 partners \cite{Barnard:2013zea} into the Sp(4) model by including of six additional fermions in the sextet, two index anti-symmetric representation $A_2$ of the gauge group.
In the nomenclature of ref.~\cite{Belyaev:2016ftv} this is model M8. The condensate breaks
the SU(6) global group of this sector to SO(6) which contains SU(3) as subgroup.
The full symmetry breaking pattern is given by
\begin{align}
\text{SU(4)} \times \text{SU(6)} \times \text{U(1)} & 
  \to \underbrace{\text{Sp(4)}}_{\text{SU(2)}_L  \times \text{U(1)}} \times 
  \underbrace{\text{SO(6)}}_{\text{SU(3)}\times \text{U(1)}} \times \text{U(1)}
\end{align}
where the U(1) factors give eventually the hypercharge.

For the holographic model one needs  the running of the gauge coupling  as outlined
in the previous section, see e.g.~\cite{Erdmenger:2020flu} for the corresponding formulae. In this model one has two condensates $\langle F F \rangle$
and $\langle A_2 A_2\rangle$ and the relevant one-loop anomalous dimensions are 
\begin{align}  \label{eq: renormalization_gherghetta_2}
\gamma_{F} =\frac{15}{8 \pi} \alpha
\quad \text{ and } \quad 
\gamma_{A_2} =  \frac{6}{2 \pi}  \alpha  \, .
\end{align}
We note, that one would expect the $A_2$ fermions to condense 
ahead of the fundamental fields since the corresponding 
critical value for $\alpha$ is smaller. 
Once the $A_2$s  condense,  SU(6) is broken to SO(6). At this 
point the $A_2$s become massive but it is unclear how quickly they decouple from the 
running of $\alpha$. This is still an open question and we refer to 
ref.~\cite{Erdmenger:2020flu} for a first discussion.

We give in table~\ref{tab:Sp4} the spectrum for various
approximations which are needed on the one hand to estimate on the on hand the
uncertainty due to the decoupling of the $A_2$ fermions and on the other hand 
for comparison with lattice data. The first column gives our results without
approximation for the various bound states: scalar $S$, axial spin-1 $A$, vector spin-1 $V$
and baryons $B$. Mesonic bound states which are either 
$A_2$ or $F$ bound states carry a corresponding index. There difference in the masses
arises as the different vacuum solutions of their respective condensate enter their
equations of motions \cite{Erdmenger:2020flu}. The baryons are $F A_2 F$ bound states.
It is still an open question how to precisely model such states within gauge/gravity duality.
We compute their masses using either the vacuum of the $F$ condensate or the one of the $A_2$
condensate as indicated by the corresponding subscript. 

For the results in the first/second column we do not/do decouple the $A_2$ fermions
at the scale where the $A_2$ condense. This clearly does not affect the masses of
the $A_2$ mesons which is the reason for us to fix the overall scale by the mass
of the vector spin-1 state $VA_2$. The decoupling of the $A_2$ implies an increase 
in the running of $\alpha$ which in turn leads to the observed increase of the masses
with $F$ index in the second column. Independent of this uncertainty we see that the
$A_2$ mesons are heavier than the $F$ mesons. Note, that in particular the mass of the
scalar $SV$ depends strongly on the decoupling behaviour of the $A_2$ fermions.
In case of the baryons, the variation 
gives a first measure of the uncertainty on the underlying mass calculation.
Last but not least we have added a NJL term to enforce an equal scale of condensation
and the corresponding results are given in the last column of this table.

\begin{table}[t] 
 \begin{center}
 \begin{tabular}{|c|ccc|cc|c|}  
\hline
     								& AdS/$Sp(4)$  	& AdS/Sp(4) 		& AdS/Sp(4) 	& lattice \cite{Bennett:2019cxd}  	& lattice \cite{Bennett:2019jzz} 	&AdS/Sp(4)\\ 
     								& no decouple 	& A2 decouple 	& quench  		& quench  									& unquench 								& + NJL \\    \hline
    $M_{V A_2}$   & 1* 		& 1* 	& 1*	& 1.000(32) & 			& 1*  \\
    $M_{V F}$ 	  & 0.61   	& 0.814 & 0.962 & 0.83(19)	& 0.83(27) 	& 1.03\\\hline
    $M_{A A_2}$   & 1.35  	& 1.35 	& 1.28 	& 1.75 (13)	& 			& 1.35\\
    $M_{A F}$ 	  &  0.938 	&1.19 	& 1.36 	&  1.32(18) & 1.34(14) 	& 1.70\\\hline
    $M_{S A_2}$   & 0.375 	& 0.375 & 1.14	& 1.65(15) & & 0.375\\
    $M_{S F}$ 	  & 0.325  	& 0.902 & 1.25 	& 1.52 (11) & 1.40(19) & 0.375\\\hline
    $M_{B A_2}$   & 1.85 	& 1.85 	& 1.86 	&	& & 1.85\\
    $M_{B F}$ 	  & 1.13 	&1.53 	& 1.79 	& 	& & 1.88\\\hline
  \end{tabular} 
  \end{center}
  \caption{ \label{tab:Sp4} AdS/$Sp(4)~4 F~ 6 A_2$. Ground state spectra for 
  various gauge/gravity models and comparison to lattice results. The subscript $A_2$ and $F$ 
  indicate the quantity in each of the two different representation sectors. 
   Note here for the unquenched lattice results, which do not include the $A_2$ fields, 
   we have normalized the $F$ vector meson mass to that of the quenched computation.}
\end{table}

Lattice studies of this model have been made in ref.~\cite{Bennett:2019cxd}
in the quenched approximation. Within gauge/gravity duality the quenching corresponds
to neglecting the hyperquark contributions to the running of the gauge coupling.
This implies a steeper running of the coupling resulting in a more compressed spectrum as can be
seen in the third column of table~\ref{tab:Sp4}.
In \cite{Bennett:2019jzz} the group followed up that work of \cite{Bennett:2019cxd} by 
unquenching the fundamental hyperquark sector using Wilson fermions. 
We show the results of both lattice studies in column 4 and 5 of this table
for direct comparison to the holographic results.  
The quenched results from both, the lattice and the holographic model, show 
considerable correlation. This provides confidence that trends as the fields 
are unquenched may be trustworthy. Consequently we would expect that if the $A_2$ 
fermions were included as unquenched fields, the $F$ sector would decrease in mass by 
about 20-40$\%$. Moreover, we also expect the scalar meson masses to be considerably 
lower than predicted by the quenched lattice computation.

A huge advantage of calculations in the gauge/gravity framework is, that one can
easily incorporate changes in the number of the underlying fermions which is much
more difficult in case of lattice calculations. We take as an example the M5 model
of ref.~\cite{Belyaev:2016ftv} which contains  \mbox{5 $A_2$} and \mbox{6 $F$} corresponding
to the following breaking pattern of the global group
\begin{align}
\text{SU(5)} \times \text{SU(6)} \times \text{U(1)} & 
  \to \underbrace{\text{SO(5)}}_{\text{SU(2)}_L  \times \text{U(1)}} \times 
  \underbrace{\text{Sp(6)}}_{\text{SU(3)}\times \text{U(1)}} \times \text{U(1)}
\end{align}
where again the U(1) factors give eventually the hypercharge. 
\begin{table}[t]
 \begin{center}
 \begin{tabular}{|c|cc|cc|cc|cc|}
    \hline 
        &  $M_{V A_2}$	& $M_{V F}$ & $M_{A A_2}$ &  $M_{A F}$ & $M_{S A_2}$ &	 $M_{S F}$
        & $M_{B A_2}$ &	$M_{B F}$ \\ \hline
    $4 F~6 A_2$ &1* & 0.618 &1.4 &0.862 &0.376 &0.348 &1.85 &1.15 \\
    $5 A_2~6F$  &1* & 0.61  &1.35 &0.938 &0.38 &0.33 & 1.85&1.13 \\
    \hline
  \end{tabular} 
 \end{center}
  \caption{\label{tab:M8vM5} For the masses in Sp(4) theories with
    two different  matter representations that can trigger chiral symmetry breaking:
    $4 F~6 A_2$ (M8 model) and $5 A_2~6F$ (M5 model).}
\end{table}
The resulting masses are given in table~\ref{tab:M8vM5}. As one would naively expect,
the results are quite similar as we have changed the fermion content only slightly.
However, note that the embedding of the SM in the corresponding global groups differ
in both models and, thus, consequently the phenomenological implications. We take
as an example the spin-1 vectors states: in case of  $4 F~6 A_2$ ($5 A_2~6F$) the ones
carrying electroweak quantum numbers will be significantly lighter (heavier) compared to the
ones carrying SU(3) quantum numbers.

\section{Phenomenological aspects}

We would like to emphasis that the implications of our results for phenomenological
investigations discussed below have to be taken with care. The results presented here so far
take are based originally on an abelian Born-Dirac-Infeld action and, thus,
they do not take into account possible multiplet structures within a given type of bound states.
Taking for example the M5 model, there is actually not one spin-1/2 state but several of them
\cite{Cacciapaglia:2021uqh} which transform as follows under the SM-gauge group
SU(3)$ \times $SU(2)$_L \times$U(1)$_Y$: $(3,2,\pm 7/6)$, $(3,2,\pm 1/6)$,
 $(3,1,\pm 2/3)$, $(8,2,\pm 1/2)$, $(8,1,0)$, $(1,2,\pm 1/2)$ and $(1,1,0)$.
The colour triplets are the usual top-partners. For a proper calculation of the
corresponding masses one has to a consider a non-abelian Born-Dirac-Infeld action as
starting which complicates the calculation quite a bit. However, we have recently
started this for some simple cases, SU(3)$\times$SU(3)$/$SU(3) \cite{Erdmenger:2023hkl}
and SU(4)$/$Sp(4) \cite{Erdmenger:2024dxf}. While these are not applicable for the M5
model directly, one can nevertheless infer some requirements needed for a successful description.

In \cite{Cacciapaglia:2021uqh} it has been shown that the M5 model contains an accidental
`baryon number' implying that either one of the baryons or the coloured pNGB transforming
as $(3,1,2/3)$, called $\pi_3$ in \cite{Cacciapaglia:2021uqh}, has to be stable. Obviously this should be the $(1,1,0)$ baryon, called $\tilde B$  in \cite{Cacciapaglia:2021uqh}, which
could then serve as a DM candidate. It might surprise at first glance that a baryon
should be lighter than a pNGB. Such a behaviour can in principle occur if the underlying
hyperquarks
have  mass terms which break the global groups explicitly as has been shown in 
\cite{Erdmenger:2020flu,Erdmenger:2024dxf}. In such a scenario $\pi_3$ would decay into $\tilde B t$ 
like the right-handed scalar
top in supersymmetric models if there is sufficient phase space. In case of smaller
mass differences,
three-body decays via an off-shell top-quark would become important, again similar to supersymmetric models
\cite{Porod:1996at,Porod:1998yp,Boehm:1999tr}. Actually, this scenario could easily be confused with supersymmetric
models at first glance as several signatures are very similar.

Assuming, that the results in table~\ref{tab:M8vM5} still give the same rough patterns,
the coloured baryons would be significantly heavier than the coloured spin-1 resonances represented
by $VF$ and $AF$, which actually represents different SU(3) states including $V_8^0$ and 
$A_3^{-2/3}$ \cite{Cacciapaglia:2024wdn}. Here the subscripts give the SU(3)
representation and the superscripts the electric charge.
 These would yield completely new
decay channels of the baryons such as
\begin{align}
\tilde g &\to A_6^{-2/3} t \to \pi_8 \pi_8 \pi^*_3 t \to 4 t + 2 \bar{t} + \tilde B   \\
T_R &\to V_8^0 t \to \pi_3 \pi^*_3 t \to 2 t+ \bar{t} + 2  \tilde B\,.
\end{align}
which to our knowledge have not yet been considered in the literature up to now.
We denoted here the $(8,1,0)$ and  $(3,1, 2/3)$ baryons by $\tilde g$ and $T_R$, 
respectively. Moreover, $\pi_8$ are strongly interacting pNGBs decaying dominantly into
$t\bar{t}$. Note also, that $\tilde B$ would lead to missing transverse momentum in the
signal. The model M5 contains also an extended pNGB sector in the electroweak sector
containing for example a doubly charged scalar $S^{++}$ which has similar quantum
numbers as the one from type II seesaw models \cite{Schechter:1980gr,Hirsch:2008gh}
and thus the same production cross section at the LHC \cite{Cacciapaglia:2022bax}. However, it
has a quite different dominant decay mode \cite{Cacciapaglia:2022bax}, namely
\begin{align}
S^{++} \to t \bar{b} W^+ \,.
\end{align}
The resulting increased hadronic activity implies that bounds on its mass are quite
weak \cite{Cacciapaglia:2022bax}. This scalar pNGB can also be produced in the decays of
top-partners. For example, is leads to an additional decay channel for the  $X_{5/3}$
baryon from the $(3,2,7/6)$ multiplet:
\begin{align}
X_{5/3} & \to W^+ t & \text{ (standard channel) } \\
X_{5/3} & \to S^{++} b \to  t b \bar{b} W^+ & \text{ (new channel).} 
\end{align}
The importance of the new channel depends of course what is the mass of the $S^{++}$ 
compared to the
baryons. We are investigating currently how important this and related decay modes are in scenarios
in which the required mass splitting of the baryons can be achieved by giving different masses
to the underlying  fermions.
We note for completeness, that exotic signatures of top-partners haven been discussed
in \cite{Bizot:2018tds,Cacciapaglia:2019zmj,Banerjee:2022xmu,Banerjee:2022izw}.

\section{Conclusions and outlook}

We have shown in this contribution that gauge/gravity duality methods are a powerful tool for obtaining
sensible estimates for masses of the bound states, both, in QCD as well as in Composite Higgs models. We have seen that reasonable
agreement with data in case of QCD and with lattice data in case of a specific Composite Higgs
model are obtained. The advantage of this framework is, that it allows for a relatively
fast computations of the spectrum,
more precisely ratio of masses. Most of the results have been obtained neglecting the fact the bound
states form multiplets of the unbroken global sub-group which can have somewhat different masses.
We have started recently to take this into account, see ref.~\cite{Erdmenger:2024dxf}, for the
case of SU(4)/Sp(4). This will be extended in different directions in the future: (i) we will take into account
fermionic bound states. (ii) We will systematically explore the models proposed in \cite{Belyaev:2016ftv}. (iii) We 
will study the impact of our results on the phenomenology of these models in a coherent global picture.

\acknowledgments
I thank  A.~Banerjee, G.~Cacciapaglia, J.~Erdmenger, N.~Evans, G.~Ferretti, Y.~Liu, T.~Flacke, J.~H.~Kim, M.~Kunkel, P.~Ko, K.~S.~Rigatos and R.~Str\"ohmer for interesting and fruitful discussions. 
I also would like to thank the organizers of the Corfu workshops for creating such a stimulating environment.
This work has been supported by DFG projects nr PO/1337-11/1 and  nr PO/1337-12/1.

\end{document}